\documentclass[nofootinbib,aps,11pt,preprintnumbers]{revtex4}
\usepackage{hyperref}
\usepackage{graphicx}
\usepackage{cancel}
\usepackage{amssymb}
\usepackage{textcomp}
\usepackage{amsmath}
\usepackage{bm}
\usepackage{times}
\usepackage{color}
\usepackage{graphics}
\usepackage{epsfig}

\begin{document}

\preprint{CALT 68-2788}

\title{\Large Dark Matter, Baryon Asymmetry, and Spontaneous $B$ and $L$ Breaking}

\author{Timothy R. Dulaney$^{1}$} \email{dulaney@theory.caltech.edu}
\author{Pavel Fileviez P\'erez$^{1,2}$} \email{fileviez@physics.wisc.edu}
\author{Mark B. Wise$^{1}$} \email{wise@theory.caltech.edu}

\affiliation{\\ \\
$^{1}$ California Institute of Technology, Pasadena, CA, 91125 USA}

\affiliation{
$^{2}$University of Wisconsin-Madison, Department of Physics \\
1150 University Avenue, Madison, WI 53706, USA}

\date{\today}
\begin{abstract}
We investigate the dark matter and the cosmological baryon asymmetry in a simple theory where baryon ($B$) and lepton ($L$) number  are local gauge symmetries that are spontaneously broken. In this model, the cold dark matter candidate is the lightest new field with baryon number and its stability is an automatic consequence of the gauge symmetry. Dark matter annihilation is either through a leptophobic gauge boson whose mass must be below a TeV or through the Higgs boson. Since the mass of the leptophobic gauge boson has to be below the TeV scale one finds that in the first scenario there is a lower bound on the elastic cross section of about $5 \times 10^{-46} \  \rm{cm}^2$.  Even though baryon number is gauged and not spontaneously broken until the weak  scale, a cosmologically acceptable baryon excess is possible. There is tension between achieving both the measured baryon excess and the dark matter density.
\end{abstract}
\maketitle

\section{Introduction}
In the LHC era, we hope to either verify the standard model  or discover the theory that describes the physics of the weak scale. One of the open issues in the standard model (SM)  is the origin of the accidental 
global symmetries, $U(1)_{B}$ and $U(1)_{L}$, where $B$ stands for baryon number and 
$L$ for the total  lepton number. At the non-renormalizable level in the SM one can find operators that violate baryon number and  lepton number. For example, $QQQl/\Lambda_{B}^2$ and $llHH/\Lambda_L$, 
where $\Lambda_B$ and $\Lambda_L$ are the scales where $B$ and $L$ are respectively broken~\cite{Weinberg:1979sa}.  Since the $QQQl/\Lambda_{B}^2$ operator gives rise to proton decay~\cite{Nath:2006ut} the cutoff of the theory has to be very large, $\Lambda_{B} > 10^{15}$ GeV.  There is no other reason that the cutoff of the SM has to be that large and so it is worth thinking  about the possibility that both $B$ and $L$ are local gauge symmetries that are spontaneously broken~\cite{FileviezPerez:2010gw} at a much lower scale (e.g., the weak scale)  and it is these gauge symmetries that prevent proton decay. 

Recently, two simple models (denoted model (1) and model (2)) where $B$ and $L$ are local gauge symmetries have been proposed~\cite{FileviezPerez:2010gw}.  In these models all anomalies are cancelled by adding a single new fermionic generation. One of the theories (model (1)) has an interesting realization of the seesaw mechanism~\cite{Minkowski:1977sc, Gell-Mann1979, Mohapatra:1979ia} for neutrino masses and they both have a natural suppression of tree-level flavor changing neutral currents in the quark and leptonic sectors due to the gauge symmetries and particle content.  In model (2), the neutrinos have Dirac masses.  In addition, for model (2), the lightest new field with baryon number is a  candidate for the cold dark matter and its stability is an automatic consequence of the gauge symmetry.  It has been shown in Ref.~\cite{FileviezPerez:2010gw} that $B$ and $L$ can be broken at the weak scale and one does not generate dangerous operators mediating proton decay.  We show how a dark matter candidate can arise in model (1).

In this article we investigate the properties of the cold dark matter candidates in the models proposed in 
Ref.~\cite{FileviezPerez:2010gw} and study the implications of spontaneous $B$ and $L$ breaking 
at the weak scale for the baryon asymmetry in the Universe.  In model (2), the dark matter candidate, $X$, which has baryon number $-2/3$  can either annihilate through the leptophobic $Z_B$ present in the theory or through the Higgs boson.  We study the constraints from the relic density and the predictions for the elastic cross section relevant for direct detection experiments. We discuss the implications of the gauging of $B$ and $L$ for baryogenesis.  There is a potential conflict between the measured baryon excess and dark matter
density.  

For model (1), we discuss the generation of a baryon excess.  We introduce a limit of the theory where $L$ is broken at a high scale but $B$ is spontaneously broken at the weak scale.  In this limit  standard leptogenesis plus a primordial excess in the field responsible for baryon number breaking can give rise to an acceptable baryon excess and dark matter density even though the baryon number gauge symmetry is not broken until the weak  scale. 

This paper is organized as follows: In Section \ref{Section2} we discuss the main features of the model. 
In Section \ref{Section3} we discuss, for model (2),  the properties of the dark matter candidate in the theory, constraints from the relic density and the predictions for the elastic cross section relevant for direct 
detection experiments. The properties of the dark matter candidate in model (1) are similar to cases already discussed in the literature (see for example \cite{LopezHonorez:2006gr} and \cite{Dolle:2009fn}). In Section \ref{Section4} we discuss the implications of the breaking of $B$ and $L$ at the weak  scale for baryogenesis.  We summarize the main results in Section \ref{Section5}.

\section{Spontaneous $B$ and $L$ Breaking} \label{Section2}
The theory proposed in Ref.~\cite{FileviezPerez:2010gw} is based on the gauge group 
$$SU(3)_C \bigotimes SU(2)_L \bigotimes U(1)_Y \bigotimes U(1)_B \bigotimes U(1)_L.$$  To fix notation, the particle content of the SM is summarized in Table~\ref{SM}.  The superscript index $(i)$ on standard model fermion fields labels the generation.  We have added three generations of right-handed neutrinos to the minimal standard model.  

\begin{table}[h]
  \centering 
  \caption{{\bf Standard Model Particle Content}}
  \label{SM}
  \begin{tabular}{|c|c|c|c|c|c|}
\hline
~~Field ~~  &~~ $SU(3)$ ~~ &~~ $SU(2)$~~ &~~ $U(1)_Y$~~ &~~ $U(1)_B$~~ &~~ $U(1)_L$ ~~ \\
\hline \hline
\rule[-4mm]{0mm}{12mm} $Q^{(i)}_L= \begin{pmatrix}  u^{(i)}_L   \\  d^{(i)}_L \end{pmatrix}$   & {\bf 3}  & {\bf 2} & ${1\over 6}$ & ${1 \over 3}$ & 0 \\
\rule[-3mm]{0mm}{11mm} $u^{(i)}_R$   & {\bf 3}  & {\bf 1} & ${2\over 3}$ & ${1 \over 3}$& 0 \\
\rule[-3mm]{0mm}{11mm} $d^{(i)}_R$   & {\bf 3}  & {\bf 1} & $-{1\over 3}$ & ${1 \over 3}$ & 0 \\
\rule[-4mm]{0mm}{12mm} $l^{(i)}_L= \begin{pmatrix}  \nu^{(i)}_L   \\  e^{(i)}_L \end{pmatrix}$   & {\bf 1}  & {\bf 2} & $-{1\over 2}$ & 0 & 1 \\
\rule[-3mm]{0mm}{11mm} $\nu^{(i)}_R$   & {\bf 1}  & {\bf 1} & 0 & 0 & 1 \\
\rule[-3mm]{0mm}{11mm} $e^{(i)}_R$   & {\bf 1}  & {\bf 1} & $-1$ & 0 & 1 \\
\rule[-4mm]{0mm}{12mm} $H = \begin{pmatrix}  H^+   \\  H^0 \end{pmatrix}$   & {\bf 1}  & {\bf 2} & ${1 \over 2}$ & 0 & 0 \\
\hline
\end{tabular}
\end{table}

When gauging $B$ and $L$, one can have two different scenarios:

\subsection{Model (1)} In this model the baryonic anomalies are cancelled by adding the new quarks $Q^{'}_L$, $u^{'}_R$ and $d^{'}_R$ which 
transform under the SM gauge group in the same way as the SM quarks but have baryon number $B=-1$.  
At the same time the leptonic anomalies are cancelled if one adds new leptons $l^{'}_L$, $\nu^{'}_R$ and $e^{'}_R$ 
with lepton number, $L = -3$.  All anomalies in the SM gauge group are cancelled since we have added one full new family. The particle content of model (1), beyond that of the SM, is summarized in the Table~\ref{BSM1}.   

\begin{table}[h!]
  \centering 
  \caption{{\bf Particle Content Beyond the SM in Model (1)}}\label{BSM1}
  \begin{tabular}{|c|c|c|c|c|c|}
\hline
~~Field ~~  &~~ $SU(3)$ ~~ &~~ $SU(2)$~~ &~~ $U(1)_Y$~~ &~~ $U(1)_B$~~ &~~ $U(1)_L$ ~~ \\
\hline \hline
\rule[-4mm]{0mm}{12mm} $Q'_L= \begin{pmatrix}  u'_L   \\  d'_L \end{pmatrix}$   & {\bf 3}  & {\bf 2} & ${1\over 6}$ & -1 & 0 \\
\rule[-3mm]{0mm}{11mm} $u'_R$   & {\bf 3}  & {\bf 1} & ${2\over 3}$ & -1 & 0 \\
\rule[-3mm]{0mm}{11mm} $d'_R$   & {\bf 3}  & {\bf 1} & $-{1\over 3}$ & -1 & 0 \\
\rule[-4mm]{0mm}{12mm} $l'_L= \begin{pmatrix}  \nu'_L  \\  e'_L \end{pmatrix}$   & {\bf 1}  & {\bf 2} & $-{1\over 2}$ & 0 & -3 \\
\rule[-3mm]{0mm}{11mm} $\nu'_R$   & {\bf 1}  & {\bf 1} & 0 & 0 & -3 \\
\rule[-3mm]{0mm}{11mm} $e'_R$   & {\bf 1}  & {\bf 1} & $-1$ & 0 & -3 \\
\rule[-3mm]{0mm}{11mm} $S_B$   & {\bf 1}  & {\bf 1} & 0 & $-{8 \over 3}$ & 0 \\
\rule[-3mm]{0mm}{11mm} $S_L$   & {\bf 1}  & {\bf 1} & 0 & 0 & 2 \\
\rule[-3mm]{0mm}{11mm} $S$   & {\bf 1}  & {\bf 1} & 0 & $-{4 \over 3}$ & $0$ \\
\rule[-4mm]{0mm}{12mm} $\phi= \begin{pmatrix}  \phi^+  \\  \phi_R^0 +  i \phi_I^0 \end{pmatrix}$   & {\bf 1}  & {\bf 2} & ${1\over 2}$ & ${4 \over 3} $& 0 \\
\hline
\end{tabular}
\end{table}

Let us discuss the main features of this scenario.

\begin{itemize}
\item \textit{Quark Sector}

In this model the masses for the new quarks are generated through the terms,
\begin{eqnarray}
-\Delta {\cal L}_{q' {\rm mass}}^{(1)}&=&Y_U^{'} \  \overline{Q^{'}_L} \  \tilde{H} \ u_R^{'} 
\ + \  Y_D^{'} \  \overline{Q^{'}_L} \  {H} \ d_R^{'}   \ + \  \rm{h.c.}.
\label{C1-quarks}
\end{eqnarray}
Here $\tilde{H}=i \sigma_2 H^*$. In order to avoid a stable colored quark, the scalar doublet $\phi$ has been added 
to mediate the decays of the fourth generation of quarks. The following terms occur in the Lagrange density
\begin{eqnarray}
-\Delta{\cal L}_{{DM}}^{(1)}&=& Y_1 \  \overline{Q_L^{'} } \  \tilde{\phi} \  u_R \ + \  Y_2 \  \overline{Q_L} \  \phi \  d'_R \ + \   \rm{h.c.}.
\label{C1-DM}
\end{eqnarray}
Here flavor indices on the Yukawa couplings $Y_i$,  and the standard model quark fields have been suppressed.
The field $\phi$ does not get a vacuum expectation value (VEV) and so there is no mass mixing between the new exotic generation of quarks and their SM counterparts.  When the real or imaginary component of $\phi$ is the lightest new particle with baryon number, it is stable.  The field $\phi$ has flavor changing couplings that cause transitions between quarks with baryon number $-1$ and the usual quarks with baryon number 1/3. However, since there is no mass mixing between these two types of quarks, integrating out the $\phi$ does not generate any tree level flavor changing neutral currents for the ordinary quarks. Those first occur at the one loop level.

\item \textit{Leptonic Sector}

The  interactions that generate masses for the new charged leptons are:
\begin{eqnarray}
-\Delta{\cal L}_{l}^{(1)}&=&  Y_E^{'} \  \overline{l^{'}_L} \  {H} \ e_R^{'}  \ + \  \rm{h.c.},
\label{C1-leptons}
\end{eqnarray}
while for the neutrinos they are
\begin{eqnarray}
-\Delta{\cal L}_{\nu}^{(1)}&=&  Y_\nu \  l  H \nu^C  \ + \  Y_\nu^{'} \  l^{'} H N \ + \
\nonumber
\\
& + &  \   \frac{\lambda_a}{2} \  \nu^C \ S_L \  \nu^C \ + \  {\lambda_b} \  \nu^C \ S_L^\dagger \ N \ + \  \rm{h.c.},
\label{C1-neutrinos}
\end{eqnarray}
where $S_L \sim (1,1,0,0,2)$ is the Higgs that breaks $U(1)_L$, generating masses for the right-handed neutrinos 
and the quark-phobic $Z^{'}_L$. We introduce the notation $\nu^C = (\nu_R)^C$ and $N= (\nu_R^{\prime})^C $.
After symmetry breaking the mass matrix for neutrinos in the left handed basis, $(\nu, \nu^{'}, N, \nu^C)$, 
is given by the eight by eight matrix
\begin{equation}
{\cal M}_{N} =
	\begin{pmatrix}
		0 
		&
		 0
		&
		0
		&
		M_D
	\\
		0
		&
		0
		&
		M_D^{'}
		&
		0	
	\\
		0
		&
		 (M_D^{'})^T
		&
		0
		&
		M_b
		\\
		M^T_D
		&
		0
		&
		M_b^T
		&
		M_a
	\end{pmatrix}.
\label{neutralino}
\end{equation}
Here, $M_D=Y_\nu v_H/\sqrt{2}$ and $M_a=\lambda_a v_L/\sqrt{2}$ are $3\times3$ matrices, 
$M_b=\lambda_b v_L^*/ \sqrt{2}$ is a $1\times 3$ matrix, $M_D^{'}=Y_\nu^{'} v_H/\sqrt{2}$ is a number 
and  $\langle S_L\rangle= v_L/{\sqrt{2}}$. Lets assume that the three right-handed neutrinos $\nu^C$ 
are the heaviest. Then, integrating them out  generates the following mass matrix for the three light-neutrinos:
\begin{equation}
{\cal M}_\nu = M_D \ M_a^{-1} \ M_D^T.
\end{equation}
In addition, a Majorana mass $M'$ for the fourth generation right handed neutrino $N,$
\begin{equation} 
M^{'}=M_b M_a^{-1} M_b^T,
\end{equation}
is generated. Furthermore, suppose that  $M^{'} << M_D^{'}$,  then the new fourth generation neutrinos $\nu^{'}$ and $N$ are quasi-Dirac with a mass equal to $M_D^{'}$. Of course we need this mass to be greater than $M_Z/2$ to be consistent with the measured $Z$-boson width. In this model we have a consistent mechanism for neutrino masses which is a particular combination of  Type I seesaw. 

\item \textit{Higgs Sector}

The minimal Higgs sector needed to have a realistic theory where $B$ and $L$ are both gauged, and have a DM candidate is composed of the SM Higgs, $H$, $S_L$, $S \sim (1,1,0,-4/3,0)$, $S_B$ and $\phi$.  $S_B$ and $S_L$ are the scalars field whose vacuum expectation values break $U(1)_B$ and $U(1)_L$, respectively, generating masses for the gauge bosons coupling to baryon number and lepton number.  
Here one introduces the scalar field $S$ in order to have a viable cold dark matter candidate. 
In this case the scalar potential of the model must contain the terms
\begin{equation}
 \mu_1 \ \left( H^\dagger \phi \right) \ S \ + \ \mu_2 \ S_B^\dagger \ S^2 \ + \ \rm{h.c.}, 
\label{C1-scalar}
\end{equation}
in order to generate the effective interaction: $ c \  (H^\dagger \phi)^2 S_B \ + \ \rm{h.c.}$, which breaks the degeneration between 
the $\phi_R^0$ and $\phi_I^0$. Here $S$ does not get the vev. Then, one of them can be a dark matter candidate and the mass splitting is given by
\begin{equation}
M_{\phi^0_R}^2-M_{\phi^0_I}^2 = \sqrt{2} \frac{v_H^2 v_B \mu_1^2 \mu_2}{M_S^4}.
\end{equation}
By adjusting the phases of the fields $S$ and $\phi$, the parameters $\mu_{1,2}$ can be made real and positive.  In this case, the imaginary part of the neutral component of $\phi$, denoted $\phi^0_I$ is the dark matter candidate.  Notice, that this DM scenario is quite similar to the case of the Inert Higgs Doublet Model since we do not have annihilation through the $Z_B$ in the non-degerate case. It is well-known that if the real and imaginary parts are degenerate in mass one cannot satisfy the bounds coming from direct detection, therefore one needs a mass splitting.  This dark matter candidate is very similar to that of the Inert Doublet Model (see, for example, \cite{LopezHonorez:2006gr} and \cite{Dolle:2009fn}).

\end{itemize}

Before concluding the discussion of model (1) one should mention that in this model local $U(1)_B$ and $U(1)_L$ are broken by the Higgs mechanism, as explained before, and one gets that in the quark sector a global symmetry (baryonic) is conserved, while in the leptonic sector the total lepton number is broken.

\subsection{Model (2)} 
In this model, the baryonic anomalies are cancelled by adding the new quarks $Q'_R$, $u'_L$ and $d'_L$ which transform under the SM gauge group the same way as the SM quarks but have opposite chirality and baryon number $B=1$.  At the same time the leptonic anomalies are cancelled if one adds new leptons $l'_R$, $\nu'_L$ and $e'_L$ with opposite chirality of their SM counterparts and with lepton number, $L = 3$. The particle content of model (2), beyond that of the SM, is summarized in the Table~\ref{BSM2}. 

\begin{table}[h!]
  \centering 
  \caption{{\bf Particle Content Beyond the SM in Model (2)}}\label{BSM2}
  \begin{tabular}{|c|c|c|c|c|c|}
\hline
~~Field ~~  &~~ $SU(3)$ ~~ &~~ $SU(2)$~~ &~~ $U(1)_Y$~~ &~~ $U(1)_B$~~ &~~ $U(1)_L$ ~~ \\
\hline \hline
\rule[-4mm]{0mm}{12mm} $Q'_R= \begin{pmatrix}  u'_R   \\  d'_R \end{pmatrix}$   & {\bf 3}  & {\bf 2} & ${1\over 6}$ & 1 & 0 \\
\rule[-3mm]{0mm}{11mm} $u'_L$   & {\bf 3}  & {\bf 1} & ${2\over 3}$ & 1 & 0 \\
\rule[-3mm]{0mm}{11mm} $d'_L$   & {\bf 3}  & {\bf 1} & $-{1\over 3}$ & 1 & 0 \\
\rule[-4mm]{0mm}{12mm} $l'_R= \begin{pmatrix}  \nu'_R   \\  e'_R \end{pmatrix}$   & {\bf 1}  & {\bf 2} & $-{1\over 2}$ & 0 & 3 \\
\rule[-3mm]{0mm}{11mm} $\nu'_L$   & {\bf 1}  & {\bf 1} & 0 & 0 & 3 \\
\rule[-3mm]{0mm}{11mm} $e'_L$   & {\bf 1}  & {\bf 1} & $-1$ & 0 & 3 \\
\rule[-3mm]{0mm}{11mm} $S_B$   & {\bf 1}  & {\bf 1} & 0 & $n_B$ & 0 \\
\rule[-3mm]{0mm}{11mm} $S_L$   & {\bf 1}  & {\bf 1} & 0 & 0 & 2 \\
\rule[-3mm]{0mm}{11mm} $S'_L$   & {\bf 1}  & {\bf 1} & 0 & 0 & $n_L$ \\
\rule[-3mm]{0mm}{11mm} $X$   & {\bf 1}  & {\bf 1} & 0 & $-{2 \over 3}$ & 0\\
\hline
\end{tabular}
\end{table}

\begin{itemize}
\item \textit{Quark Sector}

In this model the masses for the new quarks are generated through the terms,
\begin{eqnarray}
-\Delta {\cal L}_{q' {\rm mass}}^{(2)}&=&Y_U^{'} \  \overline{Q^{'}_R} \  \tilde{H} \ u_L^{'} 
\ + \  Y_D^{'} \  \overline{Q^{'}_R} \  {H} \ d_L^{'}   \ + \  \rm{h.c.}.
\end{eqnarray}
As in the previous model, one has to avoid a stable colored quark.   For this reason, we add the scalar field $X$ to mediate the decays of the fourth generation of quarks. The following terms occur in the Lagrange density
\begin{eqnarray}
-\Delta{\cal L}_{{DM}}^{(2)}&=& \lambda_Q \ X \  \overline{Q_L }\ Q_R^{'} \ + \  \lambda_U \ X \  \overline{u_R} \ u_L^{'} 
\ + \  \lambda_D \ X \  \overline{d_R }\ d_L^{'} \ + \   \rm{h.c.}.
\label{DM}
\end{eqnarray}
Here flavor indices on the Yukawa couplings $Y$,  $\lambda$ and the standard model quark fields have been suppressed.
The field $X$ does not get a vacuum expectation value (VEV) and so there is no mass mixing between the new exotic generation of quarks and their SM counterparts.  When $X$ is the lightest new particle with baryon number, it is stable.  This occurs because the model has a global $U(1)$  symmetry where the $Q'_R$, $u'_L$, $d'_L$ and $X$ get multiplied by a phase.  This $U(1)$ symmetry is an automatic consequence of the gauge symmetry and the particle content.  Notice that the new fermions have $V+A$ interactions with the W-bosons.

The field $X$ has flavor changing couplings that cause transitions between quarks with baryon number 1 and the usual quarks with baryon number 1/3. However, since there is no mass mixing between these two types of quarks, integrating out the $X$ does not generate any tree level flavor changing neutral currents for the ordinary quarks. Those first occur at the one loop level.

\item \textit{Leptonic Sector}

The  interactions for the new leptons are
\begin{eqnarray}
-\Delta{\cal L}_{l}^{(2)}&=&  Y_E^{'} \  \overline{l^{'}_R} \  {H} \ e_L^{'}  \ + \ \lambda_e \  \bar{e}_R \ S_L^\dagger e_L'  \ + \  
\nonumber \\
&+&    Y_\nu \  \overline{l_L} \  \tilde{H} \  \nu_R \ + \  Y_\nu^{'} \  \overline{l_R^{'}}  \  \tilde{H} \  \nu_L^{'}  \ + \
 \frac{\lambda_a}{2} \  \nu_R^T \ C \ S_L^{\dagger} \  \nu_R  \  
 \nonumber \\
 &+& \  {\lambda_b} \  \overline{\nu_R} \ S_L^{\dagger} \  \nu_L^{'} \ + \  \lambda_l \ \overline{l_R^{'}} \  S_L \ l_L \ + \rm{h.c.}.
\end{eqnarray}
The neutrinos are Dirac fermions with masses proportional to the vacuum expectation value of the SM Higgs boson.   Here $S_L$ must be introduced to evade the experimental constraints on heavy stable Dirac neutrino 
from dark matter direct detection and collider bounds.  In order to avoid flavor violation in the leptonic 
sector we assume that $S_L$ does not get a vacuum expectation value.

\item \textit{Higgs Sector}

The minimal Higgs sector needed to have a realistic theory where $B$ and $L$ are both gauged, and have a DM candidate is composed of the SM Higgs, $H$, $S_L$, $S'_L$, $S_B$ and $X$.  $S_B$ and $S'_L$ are the scalars field whose vacuum expectation values break $U(1)_B$ and $U(1)_L$, respectively, generating masses for the gauge bosons coupling to baryon number and lepton number.  The scalar potential of the model is given by:
\begin{eqnarray}
V_{BL}^{(2)}&=&  \sum_{\Phi_i=H,S_L,S_L^{'},S_B,X} M_{\Phi_i}^2 \Phi_i^\dagger \Phi_i 
\ + \ \sum_{\Phi_i \Phi_j} \lambda_{\Phi_i \Phi_j} \  \left(\Phi_i^\dagger \Phi_i\right) \left(\Phi_j^\dagger \Phi_j\right). 
\end{eqnarray}
In this theory one has five physical CP-even neutral Higgses  $\{H^0, S_L^0, {S'_L}^0, S_B^0, X_R^0\}$, 
and two CP-odd neutral Higgses $X_I^0$ and $S_I^0$.  Here, $X_R^0$ and $X_I^0$ have the same mass and they are cold dark matter candidates.
\end{itemize}
In this model one should notice that the local symmetries $U(1)_B$ and $U(1)_L$ are broken and after symmetry breaking one has a baryonic and leptonic global symmetries. Therefore, the proton is stable and the neutrinos are Dirac fermions. 

These are the main features of the two models that are needed to investigate the implications and/or constraints coming from cosmological observations.
\section{$X$ as a candidate for the cold dark matter in Model (2)}  \label{Section3}
As we have mentioned before, the lightest new field with baryon number, $X$,  is a cold dark matter candidate in model (2). In this section we study in detail the possible cosmological constraints and the predictions for elastic dark matter-nucleon cross section relevant for direct searches of dark matter.  Some of this material is standard and has been discussed in the literature in the context of other dark matter candidates; however, we include it for completeness.

\subsection{Constraints from the Relic Density}
There are two main scenarios for the study of the relic density. In the first case $X$ annihilates 
through the leptophobic $Z_B$ gauge boson, while in the second case $X$ annihilates through the SM Higgs.  The properties of  a SM singlet scalar dark matter candidate that annihilates through the Higgs have been investigated in many previous studies~\cite{McDonald:1993ex,Burgess:2000yq, Andreas:2008xy,Barger:2007im,He:2008qm};  however, the case of annihilation through the $Z_B$ is more specific to the model we are currently examining.

\begin{itemize}
\item  $X X^\dagger \to Z_B^* \to q \bar{q}$:

We begin by studying the case where $X$ annihilation through the baryon number gauge 
boson $Z_B$, i.e. $X X^\dagger \to Z_B^* \to q \bar{q},$ dominates the annihilation cross section. 
Here we include all the quarks that are kinematically allowed. Of course the heavy fourth generation quarks 
must be heavier than the $X$ so that they do not occur in the final state. This also limits the upper range 
of $X$ masses since the theory is not perturbatively unitary if the fourth generation Yukawa's are too large. 

The annihilation cross section through intermediate $Z_B$ in the non-relativistic limit with a quark-antiquark 
pair in the final state is given by
\begin{equation}
\sigma_{Z_B} v = \frac{2 \ g_B^4}{ 81\pi}  \frac{M_X^2}{M_{Z_B}^4} \frac{v^2}{ \left( 1 - 4{\frac{M_X^2}{ M_{Z_B}^2}}\right)^2 + {\Gamma_{Z_B}^2 \over M_{Z_B}^2}} \sum_q \Theta\left(1-{\frac{m_q}{M_X}}\right) \left(1+\left({\frac{m_q^2}{2 M_X^2}}\right)\right)
\sqrt{1-{\frac{m_q^2}{M_X^2}}} \label{Zsigmav}
\end{equation}
where $\Theta$ is the unit step function and $\Gamma_{Z_B}$ is the width of the $Z_B$.  The width of the 
leptophobic gauge boson is given by
\begin{equation}
\Gamma_{Z_B} = \sum_q {\frac{g_B^2 M_{Z_B}}{36 \pi}} 
\left(1 - {2 \frac{m_q^2}{M_{Z_B}^2}}\right) \left(1 - {4 \frac{m_q^2}{M_{Z_B}^2}}\right)^{1/2}\Theta\left(1 - {4 \frac{m_q^2}{M_{Z_B}^2}}\right).
\end{equation} 
\item $X X^\dagger \to H^* \to SM SM$:

In the case where $X$ annihilates into massive SM fields, through an intermediate $H$, 
we find that the annihilation cross section (in the non-relativistic limit) is
\begin{eqnarray}
\sigma_H v &=& \sum_f \left({\lambda_1^2N_c^f \over 4 \pi M_H^2}\right)\left({m_f \over M_H}\right)^2 {\Theta\left(1-{\frac{m_f}{M_X}}\right)\left(1- \left({m_f \over M_X}\right)^2\right)^{3/2} \over \left(1 - {4 \frac{M_X^2}{M_H^2}}\right)^2 + {\Gamma_H^2 \over M_H^2}} 
\ + \  \nonumber \\ 
\nonumber
\end{eqnarray}
\begin{eqnarray}
&+& \left({\lambda_1^2 \over 2 \pi M_H^2}\right){\Theta\left(1-{\frac{M_W}{M_X}}\right)\left(1- \left({M_W \over M_X}\right)^2\right)^{1/2} \over \left(1 - {4 \frac{M_X^2}{M_H^2}}\right)^2 + {\Gamma_H^2 \over M_H^2}} \left(1+{3 M_W^4 \over 4 M_X^4} - {M_W^2 \over M_X^2}\right) 
\ + \
\\ \nonumber 
&+& \left({\lambda_1^2 \over 4 \pi M_H^2}\right){\Theta\left(1-{\frac{M_Z}{M_X}}\right)\left(1- \left({M_Z \over M_X}\right)^2\right)^{1/2} \over \left(1 - {4 \frac{M_X^2}{M_H^2}}\right)^2 + {\Gamma_H^2 \over M_H^2}} \left(1+{3 M_Z^4 \over 4 M_X^4} - {M_Z^2 \over M_X^2}\right) 
\ + \ \\
&+& \left({\lambda_1^2 \over 64 \pi M_X^2}\right)\left(1-\left({M_H \over M_X}\right)^2\right)^{1/2} \Theta\left(1-{\frac{M_H}{M_X}}\right) \left|1 + {3 \over \left({4M_X^2 \over M_H^2} - 1\right) + i {\Gamma_H \over M_H}} \right|^2, \label{Hsigmav}
\end{eqnarray}
where $N_c^f$ is the number of colors of the particular species of fermion, $M_{W,Z}$ are the $W$ and $Z$ boson masses.
Included in the width, where kinematically allowed, is the invisible decay to dark matter.   We have ignored corrections to this formula that come from annihilation into two standard model massless gauge bosons.  For previous studies of this type of scenario see~\cite{McDonald:1993ex,Burgess:2000yq, Andreas:2008xy,Barger:2007im,He:2008qm}.  

\end{itemize}
Using these results, we are ready to compute the approximate freeze-out temperature $x_f = M_X/T_f$ assuming 
that one of the two annihilation channels dominates the annihilation of the dark matter.  Writing the thermally 
averaged annihilation cross section as $\left< \sigma v \right> = \sigma_0 (T/M_X)^n$, then the freeze-out temperature 
is given by,
\begin{eqnarray}
x_f &=& \ln\left[0.038(n+1)\left({g \over \sqrt{g_*}}\right) \ M_{Pl} M_X \sigma_0\right] 
-  \left(n + {1\over 2}\right) \ln \left[\ln \left[0.038(n+1)\left({g \over \sqrt{g_*}}\right) \ M_{Pl} \ M_X \sigma_0\right]\right]
\nonumber \\
\end{eqnarray}
where $M_{Pl}$ is the Planck mass, $g$ is the number of internal degrees of freedom and $g_*$ is the effective number of relativistic degrees of freedom evaluated around the freeze-out temperature\footnote{See, for example, \cite{Kolb:1988aj}.}.

The present day energy density of the relic dark matter particles X is given by,
\begin{equation}
\Omega_X h^2 = {1.07 \times 10^9 \over \text{GeV}} \left({(n+1) x_f^{n+1} \over \sqrt{g_*} \sigma_0 M_{Pl}}\right)
\end{equation}
where we have used the fact that $g_{*,S}(T) = g_*(T)$ in our case (all particle species have a common temperature).  The WMAP team recently gave a seven year fit  \cite{Larson:2010gs} and found the present day dark matter energy density to be $\Omega_{DM} h^2 = 0.1109 \pm 0.0056$.  

\begin{figure}[h] 
\includegraphics[height=5.25cm,angle=0]{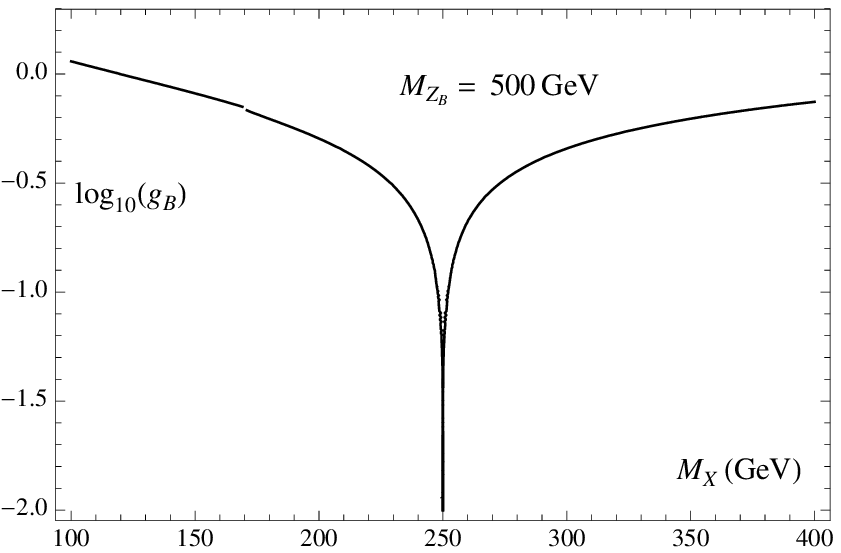}
\includegraphics[height=5.25cm,angle=0]{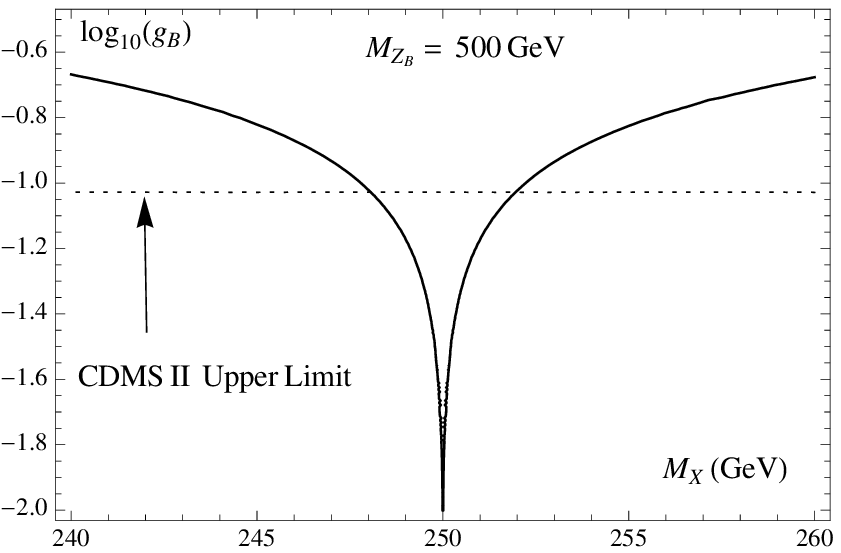}
\caption{In these figures, we plot the values of the (logarithm of the) coupling $g_B$ and dark matter mass $M_X$ 
that lead to the value of the dark matter relic abundance measured by WMAP assuming annihilation through 
intermediate $Z_B$ is dominant.  We use $M_{Z_B} = 500$ GeV for these plots.  The plot on the right 
is an enlarged version of the left plot around the region near the resonance. For dark matter masses around $250$ GeV, CDMS II excludes dark matter-nucleon elastic scattering cross sections larger than $6 \times 10^{-44} \text{cm}^2$.  The region below the dashed line is allowed by CDMS II  \cite{Ahmed:2009zw}.}
\label{ZPlots}
\end{figure}

\begin{figure}[h] 
\includegraphics[height=6cm,angle=0]{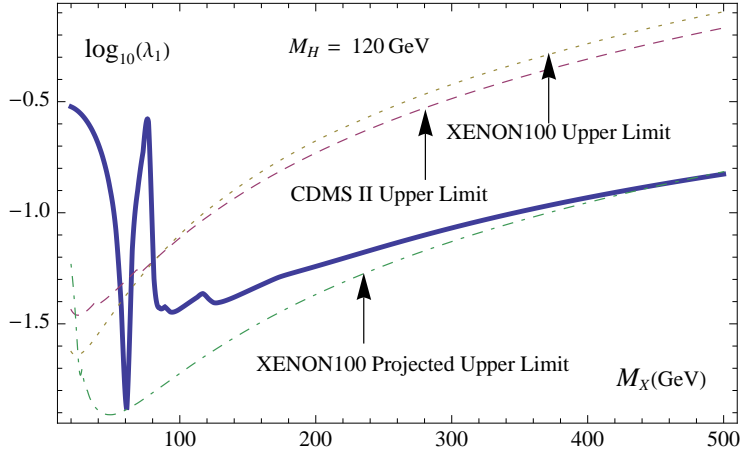}
\caption{In these figures, we plot the values of the (logarithm of the) coupling $\lambda_1$ and dark matter mass $M_X$ that lead to the value of the dark matter relic abundance measured by WMAP assuming annihilation through intermediate Higgs is dominant.  We use $M_{H} = 120$ GeV for this plots.}
\label{hPlots}
\end{figure}

Using the experimental constraints on the relic density of the cold dark matter and the annihilation cross sections calculated above, 
we plot in Figure \ref{ZPlots} (left panel) the allowed values for the gauge coupling $g_B$ and the mass of $X$ when the annihilation occurs through 
an intermediate $Z_B$ boson. Here we use as input parameter the mass of $Z_B$, $M_{Z_B} = 500$ GeV.  In order to understand the behavior of the numerical solutions close to resonance, we show the results in Figure \ref{ZPlots} (right panel), where the mass region $M_X \approx M_{Z_B}/2$ is focussed on.  
In the second scenario when the annihilation takes place through the SM Higgs boson one can display similar results.   Assuming only annihilation at tree level into SM fermions and gauge bosons for simplicity, we show in Figure \ref{hPlots} the allowed parameter space after imposing the constraints on the relic density when $M_H = 120 \ \text{GeV}$.

It is important to note that using the perturbative limit on the Yukawa couplings for the new fermions, $|Y^{'}| < 2 \sqrt{\pi}$, the masses of the new quarks, $M_{q^{'}} = Y^{'} v_H / \sqrt{2}$, are smaller than 500 GeV (since the VEV of the SM Higgs, $v_H$, is $246$ GeV).  In order to achieve the right value for the relic density, $M_X$ has to be close to the $M_{Z_B}/2$.  Hence, in the first scenario $M_{Z_B}$ must be below a TeV if $X$ annihilates primarily through the $Z_B$ and is the dark matter. This is  an acceptable kinematic range for discovery at the LHC.  Next, we  study the constraints coming from the direct detection experiments (which have already been used in the right panels of Figures~\ref{ZPlots}~and~\ref{hPlots}). 

A more precise calculation of the dark matter relic density is required when annihilation proceeds near resonance.  This is because the expansion of the annihilation cross section in terms of a polynomial in the temperature breaks down near the resonance \cite{Griest:1990kh}.  Generalizing Eq. \eqref{Zsigmav} and Eq.\eqref{Hsigmav} for general relative velocities, we determine the relic abundance near the resonance using the more precise calculation described below. The freeze-out temperature can be determined iteratively from the following equation,
\begin{equation}
x_f = \ln \left[{0.038 g M_X M_{Pl} \left<\sigma v \right> \over \sqrt{g_* x_f}}\right],
\end{equation}
where the thermally-averaged annihilation cross section is determined numerically by
\begin{equation}
\left< \sigma v \right> = {x^{3/2} \over 2 \pi^{1/2}} \int_0^\infty v^2 (\sigma v) e^{-x v^2/4} dv.
\end{equation}
The relic density is then given by,
\begin{equation}
\Omega h^2 = {1.07 \times 10^9 \over \text{GeV}} \left({1 \over J \sqrt{g_*} M_{Pl}}\right),
\end{equation}
where
\begin{equation}
J = \int_{x_f}^\infty  {\left<\sigma v \right> \over x^2} dx,
\end{equation}
takes into account the annihilations that continue to occur, but become less effective, after the freeze-out temperature.  

\begin{figure}[h] 
\includegraphics[height=5.25cm,angle=0]{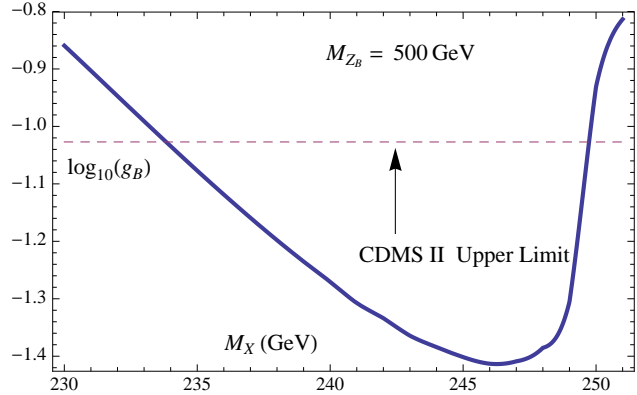}
\caption{In this figure, we plot the results of the numerical relic abundance calculation with the correct thermal averaging around the resonance.  The contour plotted shows the values of the (logarithm of the) coupling $g_B$ and dark matter mass $M_X$ that lead to the value of the dark matter relic abundance measured by WMAP assuming annihilation through an intermediate $Z_B$ is dominant.  We use $M_{Z_B} = 500$ GeV for this plot.}
\label{Zimproved}
\end{figure}

In Fig. \ref{Zimproved}, we show the contour that leads to the observed relic abundance of dark matter assuming annihilation through an intermediate $Z_B$ with mass of $500$ GeV is dominant.  After comparing this plot to the right panel in Fig. \ref{ZPlots}, it is clear that one needs to take into account the precise thermal averaging when annihilation proceeds near resonance.  The thermal averaging works to widen the contour and move the minimum below $M_{Z_B}/2$.  This is because at finite temperatures, the effective mass of the dark matter candidate is higher and therefore the minimum of the contour is shifted to lower dark matter masses.  

\begin{figure}[h] 
\includegraphics[height=5.25cm,angle=0]{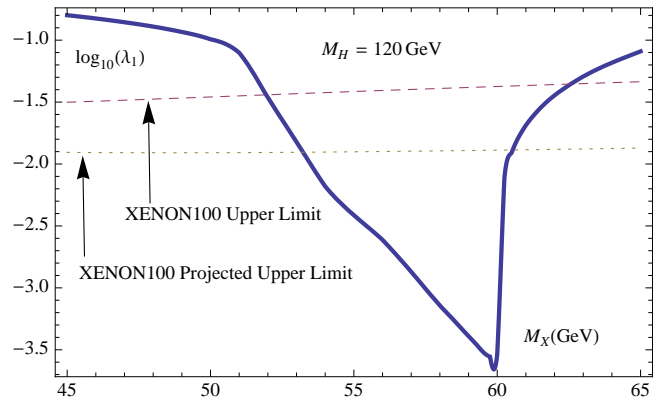}
\caption{In this figure, we plot the results of the numerical relic abundance calculation with the correct thermal averaging around the resonance.  The contour plotted shows the values of the (logarithm of the) coupling $\lambda_1$ and dark matter mass $M_X$ that lead to the value of the dark matter relic abundance measured by WMAP assuming annihilation through an intermediate Higgs is dominant and taking $M_{H} = 120$ GeV.}
\label{himproved}
\end{figure}

Similarly, in Fig. \ref{himproved}, we show the contour that leads to the observed relic abundance of dark matter assuming annihilation through an intermediate Higgs with mass of $120$ GeV is dominant. 

\subsection{Constraints from Direct Detection}
In this section we present the cross sections for elastic scattering of our dark matter candidate off of nucleons.  
These cross sections are very tightly constrained by the Cryogenic Dark Matter Search (CDMS) for dark matter masses in above approximately $100$ GeV and XENON100 for dark matter masses below approximately $100$ GeV \cite{Ahmed:2009zw,Aprile:2010um}.   

In the first scenario discussed above  we need the constraints coming from direct detection when the scattering is through  the $U(1)_B$ gauge boson.  In the non-relativistic limit, the cross section for elastic scattering of dark matter off of nucleons through an intermediate $Z_B$ is given by,
\begin{equation}
\sigma_{SI}^B = { 4 g_B^4 \over 9\pi}\left( {\mu^2 \over M_{Z_B}^4} \right)
\end{equation}
where $\mu = M_N M_X/(M_N + M_X)$ is the reduced mass of the dark matter-nucleon final state and $M_N$ is the nucleon mass.  Putting in the numbers, this cross section can be written as
\begin{equation} \label{ZDD}
\sigma_{SI}^B = (8.8 \times 10^{-40} \text{cm}^2) g_B^4 \left({500 \ \text{GeV} \over M_{Z_B}}\right)^4 \left({\mu \over 1 \ \text{GeV}}\right)^2.
\end{equation}
From the CDMS II upper limits on the spin-independent cross-section in \cite{Ahmed:2009zw}, one can conclude that if we want the correct relic abundance then $235 \ \text{GeV} \lesssim M_X \lesssim 250 \ \text{GeV}$ and $g_B \lesssim 10^{-1}$, for $M_{Z_B} \approx 500$ GeV.  For the relevant region of parameter space, see Figure \ref{Zimproved}.  

If $M_{Z_B}$ is near its ${1 \ \rm TeV}$ upper bound, the direct detection limits on the coupling $g_B$ are the weakest and the required 
range is $ 0.06 \lesssim g_B \lesssim 0.2$.  Using the plot in Fig. \ref{Zimproved}~and Eq. \eqref{ZDD}, we set a lower limit on the dark matter-nucleon scattering cross section of about $\sigma_{SI}^B \gtrsim 5 \times 10^{-46}$ cm$^2$.

For the second case when the elastic scattering of the dark matter off of nucleons is via the Higgs exchange, we need the effective coupling of the Higgs to nucleons.  For this purpose, we follow \cite{Kaplan:2000hh} and we find this effective coupling appropriate for at rest nucleon matrix element to be
\begin{equation}
{\cal L}= - {h \over v}\left(\sum_l m_l \bar{q}_l q_l + \sum_h m_h \bar{q}_h q_h\right) \rightarrow -{h \over v}\left({10 \over 27} + {17 \over 27}\hat{\chi}_+\right) M_N \left(\bar{p}p + \bar{n}n\right).
\end{equation}
Using the leading order chiral perturbation theory result in the appendix of \cite{Kaplan:2000hh} and the $\Sigma_{\pi N}$ term from \cite{Pavan:2001wz} we obtain $\hat{\chi}_+ = 0.55 \pm 0.18$ where the errors are indicative of a $30\%$ violation of $SU(3)$ flavor symmetry.   This value of $\hat{\chi}_+$ gives,
\begin{equation}
{\cal L}= -{h \over v}\left(0.72\right) M_N \left(\bar{p}p + \bar{n}n\right).
\end{equation}
With the three generations of the SM, one would have expected a number $2/9 + 7/9(0.55) = 0.65$ instead of $0.72$.  This is consistent with the $0.56 \pm 0.11$ number quoted in references \cite{Ellis:2008hf} and \cite{Farina:2009ez}.

One can use this result to compute the elastic scattering cross section,
\begin{equation}
\sigma_{SI}^H = {\lambda_1^2 \over 4 \pi} \left({10 \over 27} + {17 \over 27}\hat{\chi}_+\right)^2 \left({\mu^2 M_N^2 \over M_X^2 M_H^4}\right).
\end{equation}
Plugging in the numbers, this cross section can be written as (using $\hat{\chi}_+ = 0.55$)
\begin{equation}
\sigma_{SI}^H = (3.0 \times 10^{-41} \text{cm}^2) \lambda_1^2 \left({120 \ \text{GeV} \over M_H}\right)^4 \left({\mu \over 1 \ \text{GeV}}\right)^2\left({50 \ \text{GeV} \over M_X}\right)^2.
\end{equation}
In order to satisfy the direct detection bounds from XENON100 \cite{Aprile:2010um} for elastic scattering of dark matter off of nucleons, 
$51 \ \text{GeV} \lesssim M_X \lesssim 63 \ \text{GeV}$ with $\lambda_1 \lesssim 10^{-1.5}$, for a $120$ GeV Higgs.  This gives us a narrow region of parameter space that is not yet ruled out by the XENON100 experiment and that also leads to the correct dark matter relic abundance.   See Figure \ref{himproved} for a plot of the allowed region.  For a $120$ GeV Higgs, the dark matter-nucleon elastic cross section has a lower bound of about $\sigma_{SI}^H \gtrsim10^{-48}~ \text{cm}^2$. 

One can see from Figure \ref{hPlots} that if XENON100 reaches its projected sensitivity without detecting DM, the scenario where annihilation proceeds through the Higgs will be all but ruled out.  The only region that will be allowed from this future experiment will be the region in Figure \ref{himproved}.  For dark matter masses at the lower end of this region, the decay of the SM Higgs is dominated by the invisible decay into dark matter.  

In a more generic context, this model is different from the literature in that the dark matter mass has an upper bound (since it facilitates the decay of the fourth generation quarks and these quarks should have mass below about $500$ GeV if perturbative unitarity holds).  Most models of scalar dark matter do not have an upper limit on the dark matter mass and therefore a wider region of masses are allowed at the TeV scale. 

We need to also consider the limits direct detection experiments place on dark matter scattering off of nucleons from the interactions $\lambda X \bar{q} q'$.   To fix notation, the interactions in Eq. \eqref{DM} are
\begin{eqnarray}
-\Delta{\cal L}_{{DM}}&=& \tilde{\lambda}_Q \ X \  \bar{u}\left({1+\gamma_5 \over 2}\right) u' \ + \  \tilde{\lambda}_U \ X \  \bar{u}\left({1-\gamma_5 \over 2}\right) u' \nonumber \\
& + &  \lambda'_Q \ X \  \bar{d}\left({1+\gamma_5 \over 2}\right) d' \ + \  \lambda'_d \ X \  \bar{d}\left({1-\gamma_5 \over 2}\right) d'  \ + \   \rm{h.c.},
\label{DM2}
\end{eqnarray}
where $\{u,d\}$ ($\{u',d'\}$) are the Dirac spinors corresponding to the standard model (fourth generation) quarks and $(\tilde{\lambda}_Q)_i = U^\dagger(u,L)_i{}^j (\lambda_Q)_j$ and $(\lambda'_Q)_i = U^\dagger(d,L)_i{}^j (\lambda_Q)_j$ are the coefficients in Eq. \eqref{DM} after rotating to the mass eigenstate basis. We find the effective low energy interaction of the dark matter with the standard model quarks by integrating out the heavy fourth generation quarks.  Then, the effective interactions for non-relativistic $X$ is given by,
\begin{eqnarray}
-{\cal L}_{{eff}}  &=& \left({X^\dagger X M_X \over 2 M_{u'}^2}\right) \left(|(\tilde{\lambda}_Q)_i|^2 +|(\tilde{\lambda}_u)_i|^2  \right)(u^\dagger)^i u_i \ + \  \left({X^\dagger X \over 2 M_{u'}}\right)  \left((\tilde{\lambda}_Q)_i \ (\tilde{\lambda}_u^*)^i \ + \ (\tilde{\lambda}_Q)_i \ (\tilde{\lambda}_u^*)^i  \right)\bar{u}^i u_i 
\nonumber \\
&+& \left({X^\dagger X M_X \over 2 M_{d'}^2}\right)  \left(|(\lambda'_Q)_i|^2 + |(\lambda'_d)_i|^2  \right)(d^\dagger)^i d_i  
\ + \  \left({X^\dagger X \over 2 M_{d'}}\right)  \left((\lambda'_Q)_i \ (\lambda'_d{}^*)^i \ + \ (\lambda'_Q)_i \ (\lambda'_d{}^*)^i 
 \right)\bar{d}^i d_i 
 \nonumber \\
\end{eqnarray} 
where the flavor  index $i$ should be summed over. To get the effective interaction with nucleons, we need the nucleon matrix elements $<N| q^\dagger q |N>$ and $<N| \bar{q} q |N>$ when $q = u,d$.  We truncate the sum over flavors to the light up and down flavors.  The former simply counts the number of individual valence quarks in the nucleon and the latter matrix element is related by the coefficients $f_{Tq}$ to the former matrix elements.  This gives the effective interactions appropriate for the nucleon matrix elements,
\begin{eqnarray}
-{\cal L}_{{eff}}  & \rightarrow& \left({X^\dagger X M_X \over 2 M_{u'}^2}\right)\left(|(\tilde{\lambda}_Q)_1|^2 +|(\tilde{\lambda}_u)_1|^2  \right)(2\bar{p}p + \bar{n}n)  \nonumber \\
& + &  \left({X^\dagger X \over 2 M_{u'}}\right)\left((\tilde{\lambda}_Q)_1 \ (\tilde{\lambda}_u^*)^1 \ + \ (\tilde{\lambda}_Q)_1 \ (\tilde{\lambda}_u^*)^1  \right)f_{Tu}(2\bar{p}p + \bar{n}n)\nonumber \\
&+& \left({X^\dagger X M_X \over 2 M_{d'}^2}\right) \left(|(\lambda'_Q)_1|^2 +|(\lambda'_d)_1|^2  \right)(\bar{p}p + 2 \bar{n}n)
\nonumber \\
& + &  \left({X^\dagger X \over 2 M_{d'}}\right)  \left((\lambda'_Q)_1 \ (\lambda'_d{}^*)^1 \ + \ (\lambda'_Q)_1\ (\lambda'_d{}^*)^1  \right)f_{Td}(\bar{p}p + 2\bar{n}n). 
\end{eqnarray} 
To get an order of magnitude estimate of the size of the couplings involved, we represent the various Yukawa couplings by $\lambda$ assuming they are all the same order of magnitude.  The cross section for DM scattering off of nucleons will be small enough to evade the direct detection bounds if the Yukawa couplings, $\lambda$ are on the order of $10^{-1}$ assuming the masses of the fourth generation quarks are a few hundred GeV.  Similar constraints hold for $Y_{1,2}$ in model (1) where $\phi^0_I$ is the dark matter candidate.

\section{Cosmological Baryon  Number}  
\label{Section4}
It may be difficult to generate the observed cosmological baryon density since baryon and lepton number are gauge symmetries in the model we are considering. Here we study this issue following closely the approach of  Harvey and Turner~\cite{PhysRevD.42.3344}.  Assuming, $\mu \ll T $,  one can write the excess of particle over antiparticle as
\begin{equation} \label{boson}
\frac{n_+ - n_-}{s}=\frac{15 g}{2 \pi^2 g_*} \frac{\mu}{T},
\end{equation}
for bosons and in the case of fermions one has
\begin{equation} \label{fermion}
\frac{n_+ - n_-}{s}=\frac{15 g}{4 \pi^2 g_*} \frac{\mu}{T},
\end{equation}
where $\mu$ is the chemical potential of the particle species, 
$g$ counts the internal degrees of freedom,  $s=2 \pi^2 g_*T^3 / 45$ is the 
entropy density, and $g_*$ counts the total number of relativistic degree of 
freedom.

For each of the fields, we associate a chemical potential.  Since the chemical potential of the gluons vanishes, all colors of quarks have the same chemical potential.  Furthermore, we assume mixing between the quarks and amongst the leptons is efficient.  This reduces the number of chemical potentials to a chemical potential for each chirality of usual leptons $\{\mu_{e_L},\ \mu_{e_R},\ \mu_{\nu_L},\ \mu_{\nu_R}\}$ and quarks $\{\mu_{u_L},\ \mu_{u_R},\ \mu_{d_L},\ \mu_{d_R}\}$ as well as the fourth-generation leptons $\{\mu_{e'_L},\ \mu_{e'_R},\ \mu_{\nu'_L},\ \mu_{\nu'_R}\}$ and fourth-generation quarks $\{\mu_{u'_L},\ \mu_{u'_R},\ \mu_{d'_L},\ \mu_{d'_R}\}$.  We also have a chemical potential for each of the scalars $S_L$ and $S_B$ (denoted as $\mu_{S_L}$ and $\mu_{S_B}$, respectively), a chemical potential for $\mu_-$  for the charged field in the Higgs doublet, $\mu_0$ for the neutral Higgs field.  At temperatures above the electroweak phase transition ($T \gtrsim 300$ GeV), we  set the third component of the gauged weak isospin to zero.  This condition implies that the chemical potential for the charged W bosons vanishes and leads to the conditions
\begin{equation}
\mu_{u_L} = \mu_{d_L} ~~~~ \text{and} ~~~~  \mu_{e_L} = \mu_{\nu_L},
\end{equation}
for the SM quark and lepton fields and 
\begin{equation}
\mu_{u'_{L(R)}} = \mu_{d'_{L(R)}} ~~~~ \text{and} ~~~~ \mu_{e'_{L(R)}} = \mu_{\nu'_{L(R)}}
\end{equation}
in model 1 (2) for the fourth generation quark and lepton fields.

\subsection{Model (1)} In model (1), we also need a chemical potential for the scalar $S$, denoted $\mu_S$, a chemical potential for the charged field in the doublet $\phi$, denoted $\mu_{\phi^+}$, and a chemical potential for the neutral component of the $\phi$ doublet, denoted $\mu_{\phi}$.  Again, since the chemical potential for the charged W bosons vanishes, $\mu_{\phi} = \mu_{\phi^+}$. 

Before study the possibility to have a baryon asymmetry let us discuss the different conditions we must satisfy. 
Using Eqs. (\ref{C1-quarks}), (\ref{C1-leptons}), (\ref{C1-neutrinos}) and (\ref{C1-scalar}) one obtains
\begin{eqnarray}
\mu_0 &=& \mu_{u_R^{'}} - \mu_{u_L^{'}}, \ \ \ \mu_0 = \mu_{d_L^{'}} - \mu_{d_R^{'}}, \\
\mu_0 &=& \mu_{\nu_R} - \mu_{\nu_L}, \ \ \ \mu_0 = \mu_{\nu_R^{'}} - \mu_{\nu_L^{'}}, \\
\mu_{S_L} &=& 2 \mu_{\nu_R}, \ \ \  \mu_0 = \mu_{e_L^{'}} \ - \  \mu_{e_R^{'}}, \\
\mu_0 &=& \mu_{\phi} \ + \  \mu_{S}, \ \ \ \mu_{S_B}= 2 \mu_{S}, 
\end{eqnarray}
and
\begin{eqnarray}
\mu_{S_L} &=& - \mu_{\nu_R} - \mu_{\nu_R^{'}}.
\label{lambda-b}
\end{eqnarray}
Yukawa interactions with the Higgs boson in the SM imply the following relations,
\begin{eqnarray}
\mu_0 = \mu_{u_R} - \mu_{u_L}, ~&~&~ -\mu_0 = \mu_{d_R} - \mu_{d_L}, \label{Higgs1} \\
-\mu_0 = \mu_{e_R} - \mu_{e_L}, ~&~&~ \mu_0 = \mu_{\nu_R} - \mu_{\nu_L}. 
\label{Higgs2}
\end{eqnarray}
Now, we using these relations to write the baryon number density ($B$), lepton number density ($L$) and electric charge density ($Q$).  
We find the following expressions for these comoving number densities,
\begin{eqnarray} 
\label{C1-B1}
B^{(1)} \ &\equiv& {n_B - n_{\bar{B}} \over s} = {15 \over 4 \pi^2 g_* T}
\left( \ 12 \mu_{u_L} \ - \   12 \mu_{u_L^{'}}  -  \frac{20}{3} \mu_{S_B} \ + \ {16 \over 3} \mu_{\phi}\right), \\ 
\label{C1-L1}
L ^{(1)}\ &\equiv& {n_L - n_{\bar{L}} \over s} = {15 \over 4 \pi^2 g_* T}
\left(  \ 20 \mu_{\nu_L} \ - \ 12 \mu_{\nu_L^{'}} \ + \  8 \mu_{\phi} \ + \  4 \mu_{S_B}\right), \\ 
\label{C1-Q1}
Q^{(1)} \ &\equiv& {n_Q - n_{\bar{Q}} \over s} = {15 \over 4 \pi^2 g_* T} 
\left( \ 20 \mu_{\phi} \ + \ 9 \mu_{S_B} \ + \ 6 \mu_{u_L} \ + \ 2 \mu_{u_L^{'}} \ - \  6 \mu_{\nu_L} \ - \  2 \  \mu_{\nu'_L} \right).
\end{eqnarray}
See Tables \ref{SM} and \ref{BSM1} for the leptonic and baryonic charges. At high temperatures, each of 
the charge densities in Eqs. \eqref{C1-B1}, \eqref{C1-L1} and \eqref{C1-Q1} must vanish.  These three conditions, 
along with the sphaleron condition 
\begin{equation} 
\label{C1-sphaleron}
3(2\mu_{u_L} + \mu_{d_L} + \mu_{e_L}) + (2\mu_{u'_L} + \mu_{d'_L} + \mu_{e'_L}) 
= 9 \mu_{u_L} + 3 \mu_{\nu_L} + 3 \mu_{u_L^{'}} + \mu_{\nu_L^{'}}=0.
\end{equation}
give us four equations. Unfortunately, in the general case we do not have a symmetry which guarantees the conservation of a given number density.  We analyze the small $\lambda_b$ limit.\footnote{$\lambda_b$ must be small enough so that the mixing between the ordinary right-handed neutrinos and the fourth generation right-handed neutrino can be neglected in the early Universe, but large enough so that the fourth generation right-handed neutrino can decay.}   In this limit, we have the following approximate global symmetries:

$(B-L)_1$: $(Q_L,u_R,d_R,\phi) \to e^{i\alpha/3} (Q_L, u_R,d_R,\phi)$, 
$(l_L,e_R,\nu_R) \to e^{-i \alpha} (l_L, e_R,\nu_R)$, $S_L \to e^{-2i \alpha} S_L$, 
$S \to e^{-i \alpha/3} S$, $S_B \to e^{-2i \alpha/3} S_B$,

and 

$(B-L)_2$:  $(Q'_L, u'_R,d'_R,S) \to e^{-i\alpha} (Q'_L, u'_R, d'_R,S)$, $(l'_L,e'_R,\nu'_R) \to e^{i 3 \alpha} (l'_L, e'_R, \nu'_R)$, 
$\phi \to e^{i \alpha} \phi$, $S_B \to e^{-2i \alpha} S_B$.

Both of these approximate global symmetries are anomaly free and not-gauged.  The corresponding charge densities are given by
 \begin{equation}
 (B-L)_1=\frac{15}{4 \pi^2 g_{*} T} 
 \left(  12 \mu_{u_L} + \frac{4}{3} \mu_{\phi} - 12  \mu_{\nu_L} - 4 \mu_{S_L} - \frac{2}{3} \mu_{S} - \frac{4}{3} \mu_{S_B}\right),
\end{equation}
and
\begin{equation}
 (B-L)_2 = \frac{15}{4 \pi^2 g_{*} T} 
 \left(  - 12 \mu_{u'_L}   - 2  \mu_{S} + 12 \mu_{\nu'_L} + 2  \mu_\phi -  4 \mu_{S_B}\right).
\end{equation}
The baryon number density at late times will include the contribution of the ordinary quarks 
and the contribution from the decay of the fourth generation quarks.  In ordinary quarks we have
\begin{equation}
{1 \over 3}(3)(3) \left(\mu_{u_L} + \mu_{u_R} + \mu_{d_L} + \mu_ {d_R}\right) = 12 \mu_{u_L}.
\end{equation}
The contribution from the fourth-generation quarks ($Q' \rightarrow \phi + u_R$ and $d'_R \rightarrow \phi + Q_L$) gives
\begin{equation}
{1 \over 3}(3)\left(\mu_{u'_L} + \mu_{d'_L} + 2 \mu_{d'_R} \right) = 4 \mu_{u'_L} - 2\mu_{\phi} - \mu_{S_B}.
\end{equation}
Then, 
\begin{eqnarray}
B_f^{(1)}&=&{15 \over 4 \pi^2 g_* T} \left( 12 \mu_{u_L} + 4\mu_{u'_L}  - 2  \mu_{\phi} - \mu_{S_B} \right) \nonumber \\
&=&\frac{269}{1143} (B-L)_1 - \frac{13}{381} (B-L)_2.
\end{eqnarray}

Depending on the initial charge densities, it is possible to simultaneously explain the DM relic density and the baryon asymmetry in this scenario.  Notice that one can have leptogenesis at the high-scale if  the symmetry breaking scale for $U(1)_L$ is much larger than the electroweak scale.

\subsection{Model (2)}
In model (2), we must introduce a chemical potential for the scalar $S'_L$, denoted $\mu_{S'_L}$, and a chemical potential for the dark matter candidate $X$, denoted $\mu_X$.

The action is invariant under the transformations $S_B \rightarrow e^{i \alpha_B} S_B$ and $S'_L \rightarrow e^{i \alpha_L} S'_L$.   These automatic $U(1)$ symmetries are anomaly free, since no fermions transform under them. The symmetries are spontaneously broken by the vacuum expectation values of $S_B$ and $S'_L$, respectively;  however, at high temperatures the symmetry is restored.  We begin by assuming that in the early Universe a non-zero $S_B$ and $S'_L$ asymmetry is generated. This could occur for example from the decay of the inflaton after inflation. We examine if this can lead to the observed baryon excess.

We assume that  lepton number and baryon number are spontaneously broken at the weak scale.  In this case we have the following relations, assuming that the coupling constants $\{\lambda_a, \lambda_b, \lambda_l, \lambda_e\}$ are large enough to preserve thermal equilibrium when $T \gtrsim 300$ GeV,
\begin{eqnarray}
\mu_{S_L} &=& 2 \mu_{\nu_R}, \\
\mu_{S_L} &=& \mu_{\nu'_L} - \mu_{\nu_R}, \label{lambdab} \\
\mu_{S_L} &=& \mu_{e'_R} - \mu_{e_L}, \label{lambdal} \\
\mu_{S_L} &=& \mu_{e'_L} - \mu_{e_R}. \label{lambdae}
\end{eqnarray} 
Interactions with the Higgs boson imply the following relations,
\begin{eqnarray}
\mu_0 = \mu_{u'_L} - \mu_{u'_R}, ~&~&~ -\mu_0 = \mu_{d'_L} - \mu_{d'_R}, \label{Higgs3} \\
-\mu_0 = \mu_{e'_L} - \mu_{e'_R}, ~&~&~ \mu_0 = \mu_{\nu'_L} - \mu_{\nu'_R} \label{Higgs4}.
\end{eqnarray}
We also have the following equations relating the chemical potentials of the fourth generation quarks, ordinary quarks and the dark matter
\begin{eqnarray}
\mu_X = \mu_{u_L} - \mu_{u'_R}, ~&~&~ \mu_X = \mu_{u_R} - \mu_{u'_L} \label{X1}, \\
\mu_X = \mu_{d_L} - \mu_{d'_R}, ~&~&~ \mu_X = \mu_{d_R} - \mu_{d'_L} \label{X2},
\end{eqnarray}
assuming the couplings in Eq. \eqref{DM} are large enough that these interactions are in thermal equilibrium at high temperatures. 

We use these relations to write the baryon number density ($B$), lepton number density ($L$) and electric charge density ($Q$) in terms of $\{\mu_{u_L}, \ \mu_0, \ \mu_{S_L}, \ \mu_{S'_L}, \ \mu_{S_B}, \ \mu_X \}$.  We find the following expressions for these comoving number densities,
\begin{eqnarray} \label{B1}
B^{(2)} \ &=& {15 \over 4 \pi^2 g_* T}\left( \ 24 \mu_{u_L} \ + \ 2 n_B \mu_{S_B} \ - \ {40 \over 3} \mu_{X}\right), \\ \label{L1}
L^{(2)} \ &=&  {15 \over 4 \pi^2 g_* T}\left(  \ 28 \mu_{S_L} \ - \ 24 \mu_0 + \ 2 n_L \mu_{S'_L} \right), \\ \label{Q1}
Q^{(2)} \ &=&  {15 \over 4 \pi^2 g_* T}\left( \ 8 \mu_{u_L} \ + \ 26 \mu_0 \ - \ 6 \mu_{S_L} \ - \ 2 \mu_X \right),
\end{eqnarray}
see Tables \ref{SM} and \ref{BSM2} for the leptonic and baryonic charges.
At high temperatures, each of these charge densities in Eqs. (\eqref{B1}), (\eqref{L1}) and (\eqref{Q1}) must vanish.  
These three conditions, along with the sphaleron condition 
\begin{equation} \label{sphaleron}
3(2\mu_{u_L} + \mu_{d_L} + \mu_{e_L}) - (2\mu_{u'_R} + \mu_{d'_R} + \mu_{e'_R}) = 6\mu_{u_L}-2\mu_0+3 \mu_X=0.
\end{equation}
give us four equations and six unknowns.  We solve this system of equations in terms of the chemical potentials $\mu_{S_B}$ and $\mu_{S'_L}$ since these are the chemical potentials corresponding to the conserved charges in the transformation laws $S_B \rightarrow e^{i \alpha_B} S_B$ and $S'_L \rightarrow e^{i \alpha_L} S'_L$.  

We find that in thermal equilibrium the following relations amongst the chemical potentials,
\begin{eqnarray}
\mu_0 &=& {9 \over 8630} \left(21 n_B \mu_{S_B} - 19 n_L \mu_{S'_L}\right),~~\mu_{S_L} \ = {1 \over 8630}\left(162 n_B\mu_{S_B} - 763 n_L \mu_{S'_L}\right), \nonumber \\
\mu_X &=& {3 \over 8630}\left(247 n_B \mu_{S_B} - 18 n_L \mu_{S'_L}\right),~~\mu_{u_L}=-{3\over 3452}\left(41 n_B \mu_{S_B} + 4 n_L \mu_{S'_L}\right).
\label{solutions1}
\end{eqnarray}
Using these equilibrium relations, we  find what is called the baryon number density at late times.  The baryon number density at late times will include the contribution of the ordinary quarks and the contribution from the decay of the fourth generation quarks.  In ordinary quarks we have
\begin{equation}
{1 \over 3}(3)(3) \left(\mu_{u_L} + \mu_{u_R} + \mu_{d_L} + \mu_ {d_R}\right) = 12 \mu_{u_L}.
\end{equation}
The contribution from the fourth-generation quarks ($Q' \rightarrow X^\dagger + q$) gives
\begin{equation}
{1 \over 3}(3)\left(\mu_{u'_L} + \mu_{u'_R} + \mu_{d'_L} + \mu_{d'_R} \right) = 4 \left(\mu_{u_L} - \mu_X\right).
\end{equation}
The observed baryon excess is the sum of these two contributions and is given by
\begin{eqnarray} \label{baryonexcess1}
B^{(2)}_f &=& {15 \over 4 \pi^2 g_* T} \left(12 \mu_{u_L} + 4 \left(\mu_{u_L} - \mu_X\right)\right) \\
&=& {15 \over 4 \pi^2 g_* T}\left(4\left(4 \mu_{u_L} - \mu_X\right)\right) = -{1971\over 4315}\left( {15 n_B \over 2 \pi^2 g_*}\left({\mu_{S_B} \over T}\right)\right) - {66 \over 4315}\left( {15 n_L \over 2 \pi^2 g_*}\left({\mu_{S'_L} \over T}\right)\right)   \nonumber \\
 &\simeq& -0.46 \left( {15 n_B \over 2 \pi^2 g_*}\left({\mu_{S_B} \over T}\right)\right) - 0.02 \left( {15 n_L \over 2 \pi^2 g_*}\left({\mu_{S'_L} \over T}\right)\right).
\end{eqnarray}
Since $X$ is the cold dark matter candidate in the theory one has to check the prediction 
for the ratio between the DM density and the baryon asymmetry. 
The DM asymmetry is given by
\begin{eqnarray} 
 {n_X - n_{\bar{X}} \over s} &=& {15 \over 2 \pi^2 g_* T}\left( \mu_{X} \ - \  
 \frac{3}{2} \left( \mu_{u_L^{'}} + \mu_{d_L^{'}} + \mu_{u_R^{'}} + \mu_{d_R^{'}} \right) \right) \nonumber \\
 &=& {15 \over 2 \pi^2 g_* T}\left(  7 \mu_X - 6 \mu_{u_L} \right).
 \label{DMasymmetry}
\end{eqnarray}
Therefore,  in this case using Eq.~(\ref{solutions1}) one finds
\begin{eqnarray}
 {n_X - n_{\bar{X}} \over s} &=& {15 \over 2 \pi^2 g_* T}\left(  \frac{3516}{4315} n_B \mu_{S_B} -   \frac{99}{4315} n_L \mu_{S'_L} \right).
\end{eqnarray}
One can find an upper bound on $M_X$ using the constraint $|n_X - n_{\bar{X}}| \le n_{DM}$.  This gives the constraint
\begin{equation}
{\Omega_{DM}/M_X \over \Omega_B/ M_p} \geq  {\big| 3516 \Delta S_B -99 \Delta S'_L \big| \over 1971 \Delta S_B + 66 \Delta S'_L},
\end{equation}
where $M_p \simeq 1$ GeV is the proton mass and the observed ratio $\Omega_{DM} \simeq 5 \Omega_b$.  So in this scenario the dark matter mass must be in the range,
\begin{equation} \label{Xmasslimit}
M_X \leq M_p \left({\Omega_{DM} \over \Omega_B}\right){1971 \Delta S_B + 66 \Delta S'_L \over \big| 3516 \Delta S_B -99 \Delta S'_L \big|}.
\end{equation}
The work in Section \ref{Section3} shows that the dark matter mass must be at least $50$ GeV to obtain the correct dark matter relic density while evading direct detection limits. Depending on the initial charge densities, it is possible to simultaneously explain the DM relic density and the baryon asymmetry in this scenario.  Eq. \eqref{Xmasslimit} shows that this requires a somewhat awkward fine-tuning between the initial charge densities of the global symmetries $S_B \rightarrow e^{i \alpha_B} S_B$ and $S'_L \rightarrow e^{i \alpha_L} S'_L$.

In model (2) one can have a non-zero baryon asymmetry (even if $B$ and $L$ are broken at the low scale) if there is a primordial asymmetry in the scalar sector; however, we need physics beyond what is in model (2) to explain how this primordial asymmetry is generated.

\section{Summary}   \label{Section5}
We have investigated the cosmological aspects of two simple models, denoted (1) and (2), in which 
baryon number ($B$) and lepton number ($L$) are local gauge symmetries 
that are spontaneously broken around the weak scale.  In these models, the stability of our scalar dark matter candidate is a consequence of the gauge symmetry. 

In model (2), we studied the possible dark matter annihilation channels and found what values of the masses 
and couplings lead to the observed relic abundance of dark matter.  In the case where the s-wave annihilation through an intermediate Higgs dominates, we find that, for $M_H = 120$ GeV, in order to evade the direct detection bounds the coupling between the Higgs and the dark matter must be less than $10^{-1.5}$ and $ 51 \ \rm{GeV}\ \lesssim M_X \lesssim 63 \  \rm{GeV} $.  In the case where the p-wave annihilation through an intermediate leptophobic gauge boson dominates, we find that the coupling between the leptophobic $Z_B$ and the dark matter must be less than $0.1$ and $ 235 \ \rm{GeV}\ \lesssim M_X \lesssim 250 \  \rm{GeV} $ when $M_{Z_B}=500$ GeV.  In this case the leptophobic gauge boson has to be below the TeV scale and one finds a lower bound on the elastic cross section $\sigma_{SI}^B \gtrsim 5 \times 10^{-46} \  \rm{cm}^2$. In both cases, direct detection experiments constrain the annihilation to proceed close to resonance in order to evade direct detection and to produce the observed relic abundance of dark matter. We have shown that even though baryon number is gauged and spontaneously broken at the weak scale it is possible to generate a cosmological baryon excess.  A modest fine-tuning is needed to achieve both the measured dark matter relic abundance and baryon excess.  

In model (1), we introduced a simple mechanism to split the masses of the real of the imaginary part of the neutral component of the new scalar doublet to evade direct detection limits.  We showed that one can simultaneously achieve both the observed baryon asymmetry of the Universe and the dark matter relic abundance.  In particular,  when $L$ is broken at the high scale but $B$ is spontaneously broken at the weak scale, standard leptogenesis can be applied. 


\subsection*{Acknowledgments}
We would like to thank L. Randall for discussions and careful reading of the manuscript.
The work of P. F. P. was supported in part by the U.S. Department of Energy
contract No. DE-FG02-08ER41531 and in part by the Wisconsin Alumni
Research Foundation.  The work of M.B.W.  was supported in part by the U.S. Department of Energy under contract No. DE-FG02-92ER40701.   P.F.P. would like to thank Caltech for hospitality and support.

\bibliography{BandL}

\end{document}